\documentclass[
reprint,,
longbibliography,
amsmath, 
amssymb, 
aps, 
superscriptaddress,
prx,]
{revtex4-2}

\usepackage[ruled,vlined]{algorithm2e}
\usepackage{graphicx}
\usepackage{dcolumn}
\usepackage{bm}
\usepackage{braket} 
\usepackage{hyperref}
\usepackage{subfigure} 
\usepackage{nccmath}
\usepackage{amsthm} 
\usepackage{mathrsfs} 
\usepackage{xfrac} 
\usepackage{pifont}
\usepackage[export]{adjustbox}
\usepackage{bbm} 
\usepackage[utf8]{inputenc} 
\usepackage[normalem]{ulem}
\usepackage[dvipsnames,svgnames,table]{xcolor}
\usepackage[sort&compress]{natbib}
\usepackage{color}
\usepackage{longtable}
 
\usepackage{commath}
\usepackage{multirow}
\usepackage{array}

\definecolor{darkblue}{rgb}{0.0, 0.0, 0.4}
\definecolor{darkgreen}{rgb}{0.0, 0.5, 0.0}

\begin{document}

\title{Optimizing QAOA circuit transpilation with parity twine and SWAP network encodings}

\author{J. A. Monta\~nez-Barrera}
\altaffiliation{j.montanez-barrera@fz-juelich.de}
\affiliation{Jülich Supercomputing Centre, Institute for Advanced Simulation, Forschungszentrum Jülich, 52425 Jülich, Germany}

\author{Yanjun Ji}
\affiliation{Institute for Quantum Computing Analytics (PGI-12), Forschungszentrum Jülich, 52425 Jülich, Germany}

\author{Michael R. von Spakovsky}
\affiliation{Department of Mechanical Engineering, Virginia Tech, Blacksburg, VA 24061, USA}

\author{David E. Bernal Neira}
\affiliation{Davidson School of Chemical Engineering, Purdue University, West Lafayette, IN 47907, USA}

\author{Kristel Michielsen}
\affiliation{Jülich Supercomputing Centre, Institute for Advanced Simulation, Forschungszentrum Jülich, 52425 Jülich, Germany}
\affiliation{AIDAS, 52425 Jülich, Germany}
\affiliation{RWTH Aachen University, 52056 Aachen, Germany}

\begin{abstract}
Mapping quantum approximate optimization algorithm (QAOA) circuits with non-trivial connectivity in fixed-layout quantum platforms, such as superconducting quantum processing units (QPUs), requires a transpilation process to match the circuit to the hardware layout. This step is critical for reducing error rates on noisy QPUs. Two approaches that improve the resources required for such transpilation are the SWAP network and parity twine chains (PTC), which reduce the two-qubit gate count and circuit depth needed to represent fully connected circuits. In this work, we introduce a simulated annealing-based method that further reduces the encoding overhead of PTC and SWAP networks for QAOA circuits with non-fully connected two-qubit interactions. The method is benchmarked against various transpilers, including the Qiskit SAT mapper, demonstrating that beyond specific connectivity thresholds it achieves significant reductions in both two-qubit gate count and circuit depth. For example, for a 120-qubit QAOA instance with 25\% connectivity, our method achieves an 87\% reduction in depth and a 29\% reduction in two-qubit gates compared to the Qiskit transpiler. Finally, the practical impact of PTC encoding is validated by benchmarking QAOA on the \texttt{ibm\_fez} and \texttt{ibm\_kingston} devices, showing improved performance for systems of up to 20 qubits.

\begin{description}
	\vspace{0.2cm}
	\item[Keywords]
Parity Twine Chain, QAOA, Weighted MaxCut, ibm\_fez, Quantum Circuit Transpilation, SWAP Networks
\end{description}

\end{abstract}

\maketitle
\section{Introduction}

Certain currently available quantum processors, such as superconducting-based quantum devices, face a limited qubit connectivity, where each qubit can interact with only a few neighboring qubits. These inherent hardware constraints necessitate the process of transpilation \cite{ge2024quantum, kusyk2021survey, ji2022calibration} to execute quantum algorithms on such devices.
Transpilation transforms abstract quantum circuits into executable code by mapping logical qubits to physical qubits, addressing connectivity constraints, decomposing gates of algorithms into native hardware basis gates, and optimizing circuits to minimize depth and noise susceptibility. Beyond compilation overhead, geometric connectivity can also influence multiqubit stability and decoherence in hardware-level qubit arrays \cite{fu2024decoherence,throckmorton2022crosstalk}, further underscoring the importance of connectivity-aware implementations.

The problem of mapping logical qubits into physical qubits that need to obey connectivity constraints is known as the qubit allocation problem. This problem is identified as NP (nondeterministic polynomial-time) complete \cite{siraichi2018qubit}. Therefore, efficient heuristic transpilation that minimizes the number of gates needed is essential for executing algorithms on near-term quantum devices. Connectivity constraints are typically addressed by inserting SWAP gates, which introduce significant noise since two-qubit gates such as CNOT or CZ gates are the most dominant sources of error \cite{Pascuzzi2022}, and a single SWAP gate is typically decomposed into three CNOT gates.

A common technique for implementing circuits with dense two-qubit interactions on hardware with restricted qubit connectivity is the SWAP network strategy \cite{crooks2018performance, kivlichan2018quantum, o2019generalized, hagge2020optimal, weidenfeller2022scaling, ji2025algorithm}, which systematically inserts SWAP gates to route logical qubits through the physical topology. In this work, we focus on the linear qubit topology (1D chain), for which the SWAP-network structure and resource scaling are analytically characterized \cite{hagge2020optimal,weidenfeller2022scaling,ji2025algorithm}.
In the 1D-chain realization considered here, SWAP gates are applied between neighboring pairs in a brickwork pattern, so that the logical qubit labels advance through the chain and each logical pair becomes adjacent at some step of the schedule.
More broadly, connectivity-aware scheduling on limited-connectivity layouts such as linear arrays and ladders has been shown to yield execution times competitive with all-to-all connectivity for certain common algorithmic primitives, including the quantum Fourier transform and parity rotations \cite{holmes2020impact}.
Notably, for variational algorithms such as the quantum approximate optimization algorithm (QAOA) \cite{Farhi2014} requiring fully connected two-qubit interactions, mapping onto a 1D chain with a SWAP network achieves the shallowest circuit depth \cite{ji2025algorithm, weidenfeller2022scaling} compared to mapping onto other types of topologies. A recent novel approach is the use of parity twine chains (PTC) \cite{dreier2025connectivityaware,klaver2024}. This method has shown a reduction in the number of two-qubit gates and circuit depth needed to encode the QAOA \cite{Farhi2014} compared to the SWAP network encoding and the quantum Fourier transform \cite{Nielsen2011, coppersmith2002}.

In the case of non-all-to-all two-qubit interactions, the number of required two-qubit gates can be further reduced by relabeling the qubits such that, at a given depth of the SWAP or PTC encoding, all required interactions are realized. This problem is known as subgraph isomorphism and is NP-hard \cite{MATSUO_2023}. One approach to address it is the SATMapper \cite{MATSUO_2023, qopt_best_practices}, which reformulates the subgraph isomorphism problem as a SAT problem to determine the minimum depth at which all required interactions are satisfied. The solution is then used to obtain the corresponding qubit mapping.

In this work, PTC and SWAP encodings are compared against selected transpilation frameworks, including the Qiskit transpiler with optimization level 3 (Qiskit-T), IBM Quantum's routing pass (Qiskit-P), the IBM Quantum AI transpiler (Qiskit-AI), TKET, and the SATMapper. The application focus is the efficient implementation of QAOA circuits with non-all-to-all two-qubit interactions on connectivity-constrained near-term quantum devices. 
While optimal structured encodings such as the SWAP network are known for fully connected interactions on certain topologies, such constructions are generally unavailable for partially connected graphs. The analysis focuses on the number of two-qubit gates and the resulting circuit depth required to implement a QAOA layer for problems with different edge densities $E_d$. 

To further reduce circuit depth, we introduce an alternative to the SATMapper for the initial logical-to-physical qubit mapping, i.e., the subgraph isomorphism problem, for both PTC and SWAP encodings, which is optimized using simulated annealing (SA) \cite{sa_1983}. We choose SA because the cost function we choose is combinatorial and non-smooth: for a given qubit permutation, it evaluates how many two-qubit gates can be removed by checking, layer by layer from the outermost layers inward, whether the mapped qubit pairs correspond to native edges in the hardware coupling graph. Since this pruning-based cost has no closed-form expression and operates over a discrete space, gradient-based or exact methods are not suitable. SA is well suited for this setting, as it efficiently explores combinatorial landscapes and can escape local minima through its stochastic acceptance criterion without requiring a differentiable objective.

In addition, the impact of noise on both encodings is investigated using a depolarizing noise model within an LR-QAOA protocol \cite{Montanez-Barrera2024b} for fully connected (FC) weighted MaxCut problems (WMC). Experimental results are presented using the \texttt{ibm\_fez} QPU, using the FC LR-QAOA benchmarking protocol \cite{montanezbarrera2025} for problem sizes ranging from 5 to 25 qubits and layers from $p=3$ to 20. Finally, experiments on \texttt{ibm\_kingston} are presented for a WMC graph with  29.8\% of the total edges.

To the best of our knowledge, this work presents the first demonstration of the SWAP and PTC encodings outperforming standard transpiler methods in scenarios with reduced qubit connectivity. It also constitutes the first experimental implementation of the PTC encoding on real quantum hardware.

The paper is organized as follows. Section~\ref{Sec:Background} describes the SWAP and PTC encodings, the LR-QAOA circuit, the different transpilers, and the noise model used. Section~\ref{Sec:Methods} describes the SA strategy and the PTC decoding strategy. In Sec.~\ref{Sec:Results}, results for comparing the different transpilers and the PTC+SA and SWAP+SA encodings, the noise simulations of LR-QAOA using FC graphs, and experiments on ibm\_fez are presented. Finally, Sec.~\ref{Sec:Conclusions} provides conclusions. 

\section{Background}\label{Sec:Background}
In this section, we present a brief introduction to the LR-QAOA method employed to evaluate the SWAP and PTC encodings, the transpilers used, and the depolarizing noise model utilized. 

\subsection{LR-QAOA}
The linear ramp quantum approximate optimization algorithm (LR-QAOA) is a non-variational variant of QAOA \cite{Farhi2014}, in which the QAOA parameters follow linear annealing schedules. It is composed of alternating layers of unitary operations: a problem Hamiltonian whose ground state is unknown and a mixer Hamiltonian, whose ground state is initialized at the beginning of the evolution. In quantum optimization, the problem Hamiltonian typically encodes a combinatorial optimization problem (COP). As the system evolves, the average energy decreases, thereby amplifying high-quality solutions to the COP. 

In our simulations and experiments, the weighted MaxCut (WMC) problem is used. The WMC is defined over a weighted graph $G(V, E)$ where $V$ is the set of vertices (qubits), $E$ the set of edges (two-qubit connections needed), and each edge $(i,j) \in E$ has an associated weight $w_{ij}$. The WMC problem involves determining the partition of the vertices in the undirected graph, $G$, so that the total weight of the edges between the two sets is maximized.

Throughout the paper, the following graph metrics are used without distinction: the number of qubits $N_q = |V|$, the number of edges $N_E = |E|$, and the edge density $E_d = 2 N_E / (N_q(N_q - 1))$ which quantifies the ratio of actual edges to the maximum possible number of edges in an undirected graph. The WMC cost Hamiltonian is given by

\begin{equation}
H_C = \sum_{\{l, m\} \in E} w_{lm} \sigma_z^l \sigma_z^m, 
\end{equation}
where $\sigma_z^l$ is the Pauli-z term of qubit $l$.

$H_C$ is encoded into a parametric unitary gate given by

\begin{equation}\label{UC}
    U_C(H_C, \gamma_k)=e^{-i \gamma_k H_C},
\end{equation}
 where $\gamma_k$ is a parameter that comes from the linear ramp schedule, and $i = \sqrt{-1}$, with $i$ reserved throughout this document to represent the imaginary unit. Following this, in every second part of a layer, the mixer unitary operator, 

\begin{equation}\label{UB}
    U(H_B, \beta_k)=e^{i \beta_k H_B},
\end{equation}
is applied where $\beta_k$ is taken from the linear ramp schedule and $H_B = \sum_{l=0}^{N_q-1} \sigma_l^x$ with $\sigma_l^x$ the Pauli-x term of qubit $l$.

The three parameters that characterize LR-QAOA are $\Delta_\beta$, $\Delta_\gamma$, and the number of layers $p$. The $\beta_l$ and $\gamma_l$ parameters are given by 

\begin{equation}
\beta_l = \left(1-\frac{l}{p}\right)\Delta_\beta\ \ \mathrm{and} \ \
\gamma_l = \frac{l+1}{p}\Delta_\gamma,
\end{equation}
for $l=0, ..., p-1$. For our experimental results, $\Delta_\beta = \Delta_\gamma = \Delta_{\beta,\gamma}$ are used where $\Delta_{\beta,\gamma} = 0.63$ for $N_q \le 15$ and $\Delta_{\beta,\gamma} = 0.3$ for $N_q > 15$. These values are taken from the experiments of the benchmarking of LR-QAOA (see \cite{montanezbarrera2025}) and allow for comparison with previous results in the benchmarking using the SWAP strategy. 

The approximation ratio $r$ and the success probability $p_{gs}$ are used as metrics of the performance of the WMC using the encodings. The approximation ratio is given by 

\begin{equation}
r = \frac{\sum_{m=1}^{n} C(x_m)/n}{C(x^*)},
\label{eq:r}
\end{equation}
where
\begin{equation}
C(x) = \sum_{k,l>k}^{N_q} w_{kl}(2x_kx_l - x_k - x_l),
\end{equation}
and $n$ is the number of samples, $x_m$ the $m$\textsuperscript{th} bitstring obtained from LR-QAOA on a given QPU, $C(x)$ the cost function of WMC, $x^*$ the optimal bitstring, $C(x^*)$ the maximum cut, $w_{kl}$ the weight of the edge between nodes $k$ and $l$, and $x_k \in \{0,1\}$ the $k$\textsuperscript{th} position of the $x$ bitstring. The probability of success, $p_{gs} = \textrm{probability}(x^*)$, is calculated as the number of bitstrings $x^*$ obtained over the total number of sampled bitstrings.

\subsection{SWAP strategy}

In the linear nearest-neighbor setting considered here, the SWAP network defines a structured routing schedule in which alternating layers of nearest-neighbor SWAP gates move logical qubits through the chain and create the adjacencies required for two-qubit interactions. This 1D construction admits explicit analytical resource counts.

The optimal SWAP network for \( N_q \) qubits has been proven \cite{weidenfeller2022scaling, ji2025algorithm, hagge2020optimal} to require \( N_q-2 \) total swap layers to achieve all-to-all connected two-qubit gates.

To implement $U_C(\gamma) = e^{-i \gamma H_C}$ on 1D hardware, a SWAP network that covers all $\binom{n}{2}$ qubit pairs in $n$ layers via odd-even transposition is used. Let $\sigma^{(k)}$ be the logical-to-physical qubit permutation after $k$ swap rounds with $\sigma^{(0)} = \mathrm{id}$. At layer $k \in \{0,\ldots,N_q{-}1\}$, the nearest-neighbor pairs are
\begin{equation}
    L_k = \bigl\{\bigl(\sigma^{(k)}_j,\,\sigma^{(k)}_{j+1}\bigr) 
    : j \equiv k \!\!\!\pmod{2}\bigr\},
\end{equation}
with $0 \le j < N_q{-}1$. After each layer, the permutation updates by swapping 
adjacent elements at the active positions such that
\begin{equation}
    \sigma^{(k+1)}_j = 
    \begin{cases} 
        \sigma^{(k)}_{j+1} & \text{if } j \equiv k \pmod{2} \text{ and } j < N_q{-}1,\\ 
        \sigma^{(k)}_{j-1} & \text{if } j{-}1 \equiv k \pmod{2} \text{ and } j > 0,\\ 
        \sigma^{(k)}_j     & \text{otherwise}.
    \end{cases}
\end{equation}

For the standard QAOA setting, namely, a fully connected cost layer generated by pairwise $ZZ$ interactions together with the usual single-qubit $X$ mixer, the routed implementation of one cost layer on a 1D chain can be described using the rotation layer
\begin{equation}
    R_{L_k}(\gamma) = \prod_{\substack{(l,m) \in L_k \\ (l,m) \in E}} 
    \!\!\exp\!\Bigl({-}i\,\gamma\, w_{lm}\, Z_l Z_m\Bigr).
\end{equation}
The full cost unitary is
\begin{equation}\label{eq:swap_network}
    U_C(\gamma) = R_{L_{N_q-1}} 
    \!\prod_{k=n-2}^{1}\! \bigl(\mathrm{SWAP}_{L_k} \cdot R_{L_k}\bigr)
    \cdot R_{L_0}.
\end{equation}
In the middle layers ($1 \le k \le N_q{-}2$), each SWAP and $R_{ZZ}$ pair combines into three CNOTs, i.e.,
\begin{align}\label{eq:fused_swap}
    &\mathrm{SWAP}_{j,j+1} \cdot R_{ZZ}(\theta)_{j,j+1} = \nonumber\\
    &\quad \mathrm{CX}_{j,j+1}\, R_Z(\theta)_{j+1}\, 
    \mathrm{CX}_{j+1,j}\, \mathrm{CX}_{j,j+1}.
\end{align}
When $(l,m)\notin E$, the $R_Z$ is omitted, 
leaving a bare SWAP. The first and last layers use only $R_{ZZ}$ gates without SWAPs. This approach requires $N_{g} = \frac{3}{2}N_q^2 -\frac{5}{2}N_q + 1$ two-qubit gates and has two-qubit depth $d = 3N_q - 2$.
Here, $N_g$ and $d$ refer to the two-qubit gate count and two-qubit gate depth. The single-qubit $X$ mixer does not contribute to these two-qubit resource counts. 

Additionally, a recent study \cite{ji2025algorithm} reports that for fully connected two-qubit gates, the optimal SWAP strategy requires \( N_q-2 \) swap layers on a T-shaped topology and \( N_q-1 \) swap layers on an H-shaped topology. While the T- and H-shaped subtopologies offer enhanced qubit connectivity, thereby reducing the number of required SWAP gates, the increased connectivity comes at the expense of increased circuit depth compared to the linear topology \cite{ji2025algorithm}. However, irrespective of the chosen spatial topology, mirror symmetry of swap layers enables efficient construction of higher QAOA depths. The optimized swap sequence at QAOA depth one can be extended to depth $p+1$ by alternating original and mirrored swap layers, yielding a repetitive structure with period two and enabling depth scaling without additional computational effort \cite{ji2025algorithm}. Beyond addressing connectivity constraints, SWAP networks have also been employed to optimize algorithmic performance of QAOA~\cite{hashim22optimized}, enable effective error mitigation strategies~\cite{ji2023improving, ji2024synergistic}, and enhance pulse-level efficiency~\cite{ji2023optimizing}.

For Hamiltonians with partially connected two-qubit interactions, SWAP layers derived from fully connected cases can still satisfy hardware connectivity constraints by substituting the ZZ-SWAP gates associated with absent interactions with conventional SWAP gates. However, the residual SWAP gates resulting from these missing interactions can significantly amplify noise-induced errors. A practical approach to mitigate this problem is to optimize the initial qubit mapping or qubit ordering \cite{ji2025algorithm, ji2023improving}, strategically pushing all residual SWAP gates towards the end of the circuit. Subsequently, these terminal SWAP gates can be eliminated by appropriately adjusting the measurement order. However, finding the optimal initial qubit order among the $n!/2$ distinct permutations is challenging. An alternative heuristic approach, which involves exploring a subset of initial qubit orderings and selecting the permutation that minimizes the number of CNOT gates, has been proposed~\cite{ji2023improving}, which, nevertheless, becomes inefficient for large qubit systems. To overcome this, we propose a scalable, SA-based heuristic algorithm that optimizes the initial qubit order efficiently while maintaining high-quality solutions even for large circuit sizes, which we will discuss in Sec.~\ref{subsec:sa_for_qu}.

\subsection{Parity Twine Chains}

\begin{figure}[tb]
    \centering
    \includegraphics[width=1\linewidth]{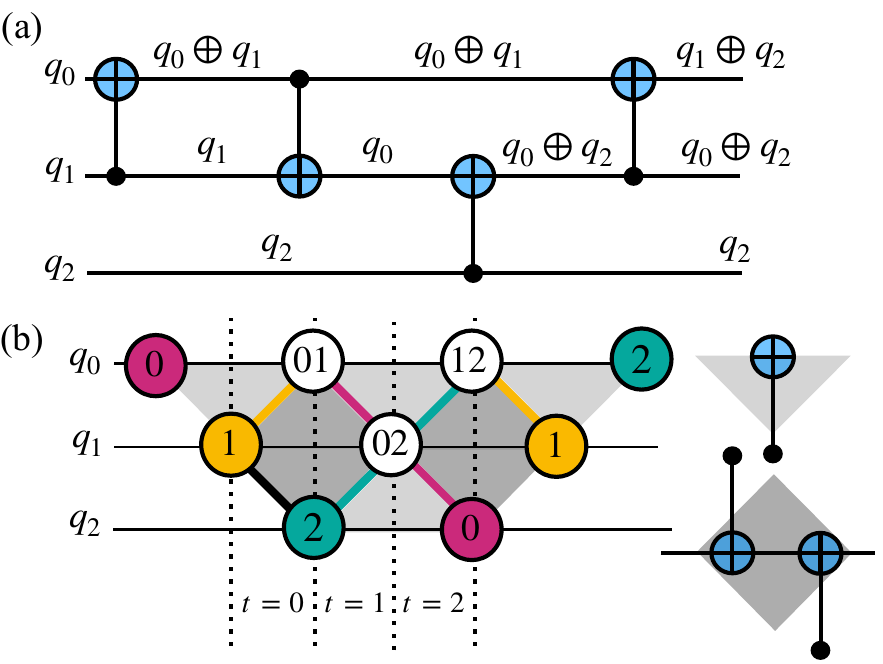}
    \caption{Schematic representation of the PTC encoding. (a) Circuit model to get different parities. (b) Graphic representation of the PTC in a 1D chain. Using this diagram, the equivalent building blocks for the CNOT gates are shown at the right. }
    \label{fig:PTC}
\end{figure}

The PTC method \cite{dreier2025connectivityaware} is a generalization of temporal parity encoding \cite{klaver2024}. In this approach, multi-qubit $Z$ information is encoded in single physical qubits, enabling local manipulation of non-local interactions with single-qubit gates. Each physical qubit~$a$ is assigned a label $P_a^{t} \subseteq \{0, 1, \ldots, N_q{-}1\}$ that tracks which logical qubits it encodes at time step~$t$, starting from $\mathcal{P}_a^{(0)} = \{a\}$ in the computational basis. A $\mathrm{CNOT}_{control \to target}$ gate updates the target label via the symmetric difference between sets $\triangle$, i.e.,
\begin{equation}\label{eq:symmetric_difference}
    \mathcal{P}_{target}^{(t)} = \mathcal{P}_{control}^{(t-1)} \triangle \mathcal{P}_{target}^{(t-1)},
\end{equation}
while leaving the control label unchanged.
Physical qubits with $|\mathcal{P}_a^{(t)}| = 1$ are called \emph{base qubits}, and those with $|\mathcal{P}_a^{(t)}| \geq 2$ are \emph{parity qubits}. At any time $t$, a physical $Z_a$ rotation acts as the logical operation $\prod_{j \in \mathcal{P}_a^{(t)}} \bar{Z}_j$, so that a single-qubit $R_z$ gate on a parity qubit effectively implements an interaction term between all logical qubits in its label~\cite{klaver2024}. This enables the execution of multi-qubit operations locally, by operating on a single physical qubit. Fig.~\ref{fig:PTC}(a) shows the circuit used to generate parities among all the qubits in a 3-qubit system. Where \(q_0=\alpha_0|0\rangle + \nu_0 |1\rangle\) and \(|0\rangle \oplus |0\rangle = |0\rangle\), \(|0\rangle \oplus |1\rangle = |1\rangle\), \(|1\rangle \oplus |0\rangle = |1\rangle\), and \(|1\rangle \oplus |1\rangle = |0\rangle\). First, a CNOT gate between physical qubits 1 and 0 generates the parity $q_0 \oplus q_1$ in physical qubit 0. The parity $q_0 \oplus q_2$ in physical qubit 1 is then created by performing CNOT gates on qubit pairs (0, 1) and (2, 1). Finally, the parity $q_1 \oplus q_2$ in physical qubit 0 is created by applying a CNOT gate between qubits 1 and 0. The key advantage of this method arises when applying two-qubit operations such as the ZZ interaction, which can now be implemented locally as a single-qubit RZ rotation on the qubit encoding the corresponding parity.

Figure~\ref{fig:PTC}(b) provides a visual representation of the PTC scheme, showing the circuit from Fig.~\ref{fig:PTC}(a) in a more abstract form. In this schematic, light-gray triangles denote two-qubit gates, while gray rhombi indicate two concurrent two-qubit gates acting on the same physical qubit. The dotted lines mark distinct time steps during which physical qubits have the labels falling in between the lines. For instance, at $t=1$ information of $q_0 q_1$ is in physical qubit 0, $q_0q_2$ in physical qubit 1, and qubit 2 in physical qubit 2. The possible two-qubit parities are generated using only a single rhombus and two triangles.

The labels also determine how logical $\bar{X}_j$ operations (required for the QAOA mixer $U_X(\beta) = \prod_j e^{-i\beta \bar{X}_j}$) are implemented physically. A logical $\bar{X}_j$ is realized by flipping all physical qubits whose label contains index~$j$, i.e., $\bar{X}_j = \prod_{b \in \mathcal{L}_j^{t}} X_b$, where the \emph{logical line} $\mathcal{L}_j^{t}$ collects the indices of all physical qubits encoding $\bar{Z}_j$ information at time~$t$~\cite{klaver2024}.
Crucially, on a linear architecture, the logical lines at any given time span at most two neighbouring qubits, and a spanning line always has at least one localized logical $X$ operator (logical line of length~$1$) on a boundary. As a result, each $e^{-i\beta \bar{X}_j}$ rotation can be implemented with at most one additional CNOT pair, so that the full mixer $U_X(\beta)$ requires a constant two-qubit gate depth of~$4$ and at most $2N_q - 2$ CNOT gates.

Using the PTC encoding, the number of two-qubit gates required to implement a single QAOA layer on a 1D chain is given by $N_g = N_q^2 - 1$, and the depth by $d = 2N_q + 2$. This is a reduction of $\Delta N_g = N_g^{\text{SWAP}} - N_g^{\text{PTC}} =\frac{1}{2}N_q^2 - \frac{5}{2}N_q + 2$ two-qubit gates compared to the SWAP network method. Note that at $t=1$, the physical qubits have the information of $q_1q_2$, $q_0q_2$, and $q_0$. Instead of using extra gates to move the physical qubits into the logical qubits, one can measure at $t=1$ and use a decoding method to extract the information of the logical qubits from the parity basis. We describe the procedure we use in Sec.~\ref{Sec:decoding}.

\subsection{Transpilation}\label{Sec:transpilation}

To evaluate the effectiveness of transpilation techniques across different $E_d$, the Qiskit \cite{qiskit2025} default transpiler (Qiskit-T), the PassManager new Qiskit functionality (Qiskit-P) \cite{qiskit_transpiler_api}, the AI Routing Pass from IBM Quantum \cite{IBM_AI_Routing}, (Qiskit-AI), and the TKET transpiler \cite{tket_docs} are used.

Qiskit is an open-source quantum computing software development kit (SDK) developed by IBM, designed for building and transpiling quantum circuits. For transpilation, Qiskit uses the VF2Postlayout algorithm \cite{Nation_2023} that consists of finding the best isomorphic subgraphs to the input circuit using a heuristic objective function. In Qiskit, there are four levels of transpiling optimization (from 0 to 3), with 3 being the level where the most effort in the algorithm occurs to find the layout with the lowest number of gates. From a recent study, it is also found to be the most performant transpiler available \cite{nation2025benchmarkingperformancequantumcomputing}.

Recently, IBM Quantum introduced the \texttt{PassManager} and the AI Routing Pass as part of the transpiler passes designed to optimize quantum circuit execution on IBM QPUs. Qiskit-P allows implementation of the SWAP network strategy directly in the transpilation sequence, and Qiskit-AI uses machine-learning techniques to predict the best qubit mapping and swap strategy, reducing circuit depth and two-qubit gate count. Qiskit (v1.4.2) is employed for quantum circuit construction and transpilation.

TKET is a quantum SDK by Quantinuum that allows users to build, compile, and run quantum circuits on multiple backends. It offers tools for circuit construction, optimization, and execution. TKET includes features for gate reduction and qubit routing, supporting various quantum devices and simulators.
In this study, the \texttt{GraphPlacement} method for initial qubit mapping is utilized, and the default routing approach, \texttt{RoutingPass}, is employed. Subsequently, TKET's built-in optimization passes, specifically \texttt{DecomposeBRIDGE}, \texttt{DecomposeSWAPtoCX}, and \texttt{RemoveRedundancies}, are applied. Finally, Qiskit's transpiler with optimization level 3 is invoked to decompose gates into the backend-specific basis gate set. TKET (v2.1.0) is utilized for quantum circuit construction and transpilation.

\subsection{SATMapper}

The SATMapper \cite{MATSUO_2023} formulates the QAOA initial mapping problem as a SAT encoding of a subgraph isomorphism between the problem graph and the hardware graph under a fixed SWAP-layer schedule. Boolean variables represent the assignment of logical qubits to physical qubits across layers, and a SAT solver is used (via a binary search over an increasing number of SWAP layers) to find a mapping that allows all two-qubit interactions of the original problem to be executed. While effective, the encoding introduces $\mathcal{O}(n^2)$ variables, $\mathcal{O}(n^3)$ constraints, and $\mathcal{O}(\log n)$ iterations, leading to rapidly growing memory and time usage, which makes the approach infeasible for large graphs and requires considerable runtime due to solving $\mathcal{O}(\log n)$ SAT instances.

To address this, the authors introduce a timeout for the SAT solver at each depth, namely, if no solution is found within the given time, the SWAP depth is increased by one and the process is repeated until a feasible encoding is obtained. For larger graphs, they further employ a clustering heuristic that partitions the problem graph into smaller subgraphs, solves the SAT mapping locally, and then combines the results, reducing memory requirements and enabling scalability, at the cost of reduced solution quality. We compare the SA approach against this method for a 200-qubit problem.

\subsection{Noise model}

On current superconducting QPUs operating on physical and unencoded qubits, two-qubit entangling gates are typically the dominant source of instruction-level noise \cite{Pascuzzi2022}. This regime is distinct from fault-tolerant architectures, where the dominant resource bottlenecks can shift to logical non-Clifford operations. In this work, a simplified depolarizing noise model is, therefore, adopted acting only on the two-qubit gates.

This channel is given by

\begin{equation}\label{Eq:depolarizing}
\mathcal{E}[\rho] = (1 - \varepsilon_g) \rho + \varepsilon_g \frac{I}{4},
\end{equation}
where $\varepsilon_g$ is the depolarizing error, $I$ is a $4 \times 4$ identity matrix, and $\rho$ is the density state operator of the two-qubit system. In general, the action of a two-qubit gate on a density state operator that represents a quantum circuit can be expressed by 

\begin{equation}
\mathcal{E}_{kl}[\rho] = (1- \varepsilon_g)U_{2Q}^{kl}\rho U_{2Q}^{kl} + \frac{\varepsilon_g}{4} \mathrm{Tr}_{kl}(\rho) \otimes I,    
\end{equation}
where $\mathcal{E}_{kl}$ is the channel acting on $\rho$, $\mathrm{Tr}_{kl}$ is the partial trace over qubits $k$ and $l$, and $U_{2Q}^{kl}$ is the two-qubit unitary gate. For simplicity, it is assumed that $\varepsilon_g$ is the same for all the two-qubit gates and models how it affects the LR-QAOA output for the PTC and SWAP encoding methods.

\begin{figure*}[htb]
    \centering
    \includegraphics[width=1\linewidth]{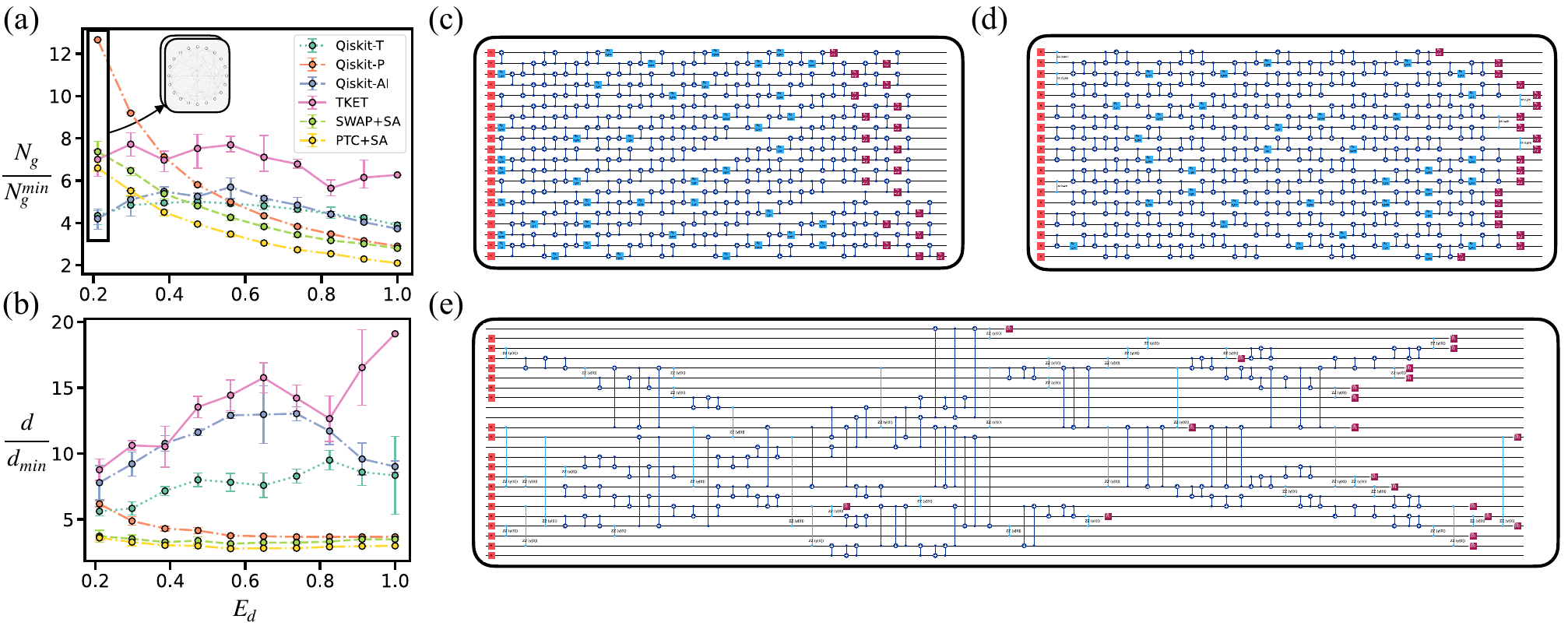}
    \caption{Resources needed to implement a layer of QAOA using the Qiskit transpilation method level 3 on a heavy-hex layout (Qiskit-T), the Qiskit PassManager (Qiskit-P), IBM Quantum AI pass (Qiskit-AI), the TKET transpiler (TKET), and the SWAP+SA and PTC+SA strategies on a 1D-chain graph. (a) Ratio of two-qubit gates used by each method to the minimum required vs. the edge density of the graph for random problems with $N_q=20$ qubits. The inset graphs represent the problem with $E_d = 0.211$. (b) Circuit depth for the same cases of (a). LR-QAOA circuit for one layer of (c) PTC+SA encoding, (d) SWAP+SA encoding, and (e) Qiskit-T encoding.}
    \label{fig:transpilation}
\end{figure*}

\section{Methods}\label{Sec:Methods}
In this section, we present the SA strategy we proposed to reduce two-qubit gate count for the SWAP-network and PTC strategies.

\subsection{Simulated annealing for qubit order optimization\label{subsec:sa_for_qu}}

The performance for circuits with non-all-to-all connected two-qubit interactions can be significantly improved by optimizing the initial qubit order in both PTC and SWAP. Here, we introduce a SA \cite{bertsimas1993simulated} heuristic algorithm to this end and show results for the SWAP and PTC strategy by minimizing a cost function that depends on the initial qubit order and estimates the total number of two-qubit operations required to realize the target interactions.
We evaluate the cost by counting the number of two-qubit interactions in the transformed graph, representing edges absent from the fully connected graph, that can be omitted at the end of the QAOA layer. These interactions, which include unnecessary ZZ and CNOT operations, can be safely removed. To optimize this, we iteratively explore the space of initial qubit labeling by randomly swapping pairs of qubit labels.

\begin{figure}[tb]
    \centering
    \includegraphics[width=.9\linewidth]{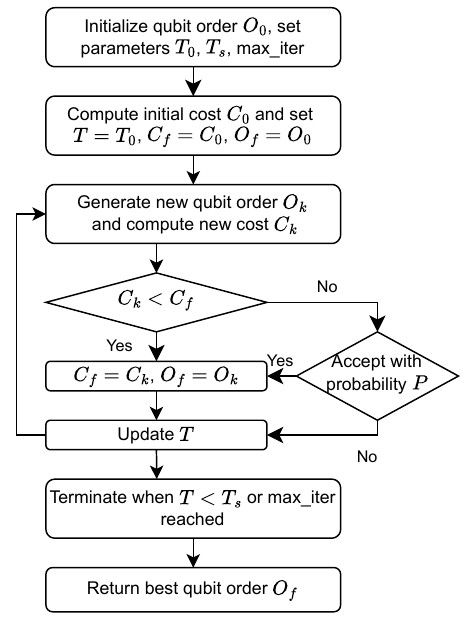}
    \caption{Flowchart of the simulated annealing algorithm for optimizing initial qubit order in SWAP-network and PTC encodings.}
    \label{fig:sa_diagram}
\end{figure}

Figure~\ref{fig:sa_diagram} illustrates the workflow of the SA algorithm employed to optimize the initial qubit order in SWAP networks and PTC. The algorithm begins with an initial permutation \(O_0\), initial temperature \(T=T_0\), stopping temperature \(T_s\), and maximum iterations max\_iter, and iteratively explores neighboring permutations by randomly swapping pairs of qubit labels. Each new configuration is evaluated using a cost function that quantifies the number of two-qubit interactions not present in the problem and that are created at the end of the PTC and SWAP encodings and, therefore, can be removed. If the new order yields a lower cost, it is accepted; otherwise, it may still be accepted with a probability \(P=\exp(-\Delta / T)\), where \(\Delta\) is the difference between the new and current costs. The temperature \(T\) is reduced by the cooling rate $\delta$ such that \(T = \delta T\), and the process continues until reaching \(T_s\) or max\_iter. The best qubit order \(O_f\) encountered is returned. The parameters used in this work are \(T_0=0.01\), \(\delta=0.999\), and max\_iter=100{,}000.

\begin{figure}[tb]
    \centering
    \includegraphics[width=.9\linewidth]{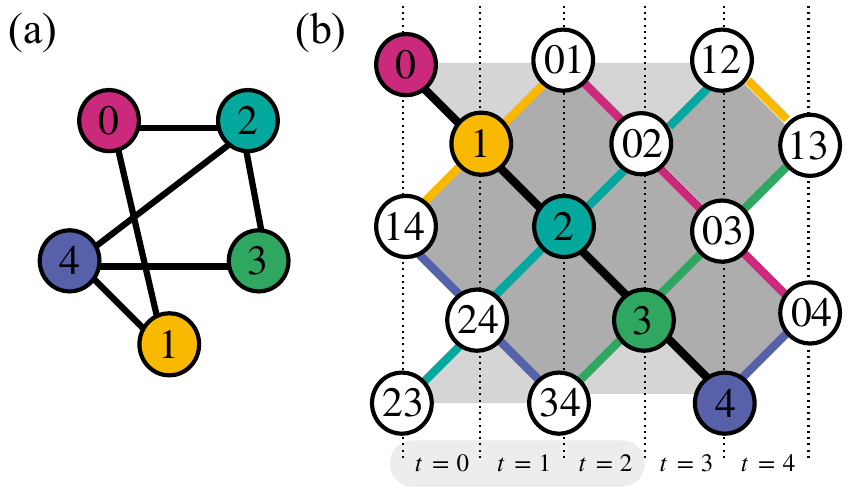}
    \caption{QAOA problem encoding with a reduced depth for non-fully connected graphs. (a) The problem graph to be encoded in a QAOA circuit. (b) PTC circuit to generate all pairs of qubit interactions.}
    \label{fig:sa_explained}
\end{figure}

In Figure~\ref{fig:sa_explained}, we illustrate a problem graph that can be mapped to a QAOA circuit with a reduced depth. The graph in Fig.~\ref{fig:sa_explained}(a) is not fully connected, with the missing edges (0, 3), (0, 4), (1, 2), and (1, 3). When using the PTC encoding, these connections are implemented at time steps \(t=3\) and \(t=4\). However, since these connections are not required in the original problem graph, the circuit can be truncated at 
\(t=2\) for the first QAOA layer. Subsequent layers can then proceed in the reverse direction, effectively alternating forward and backward passes as the QAOA \(p\) increases. The intuition behind the SA optimization for PTC and SWAP is similar: a qubit relabeling is sought such that an isomorphic version of the problem graph fits within a reduced circuit depth. This allows one to discard unnecessary time steps in the encoding, thereby minimizing resource usage. Additionally, although this work focuses on linear topology, the proposed SA-based optimization method extends effectively to QAOA implementations on T- and H-shaped topologies \cite{ji2025algorithm}, providing similar resource efficiency gains.

Extension to the general topologies of current IBM QPUs, such as the heavy-hex lattice, first requires establishing the corresponding structured encoding schedules for those topologies, which is an open problem and is, thus, left for future work.

\subsection{PTC decoding strategy}\label{Sec:decoding}
In the PTC encoding, after $p$ QAOA layers, the physical qubits are left in a 
parity basis rather than the logical (computational) basis. Specifically, at the 
end of the circuit, each physical qubit $i$ encodes either a single logical qubit 
$q_a$ or the parity of two logical qubits $q_a \oplus q_b$. The key insight is 
that instead of appending additional CNOT gates to transform back into the 
computational basis (which would increase the circuit depth), one can measure 
directly in the parity basis and perform a classical postprocessing step to 
recover the logical bitstring.

The parity map $\mathcal{P}$ assigns to each physical qubit $a$ a set $\mathcal{P}_a^{t} \subseteq \{0, 1, \ldots, N_q - 1\}$ with $|\mathcal{P}_a^t| \in \{1, 2\}$. When $|\mathcal{P}_a^t| = 1$, the physical 
qubit directly encodes logical qubit $q_x$, and its measurement outcome $m_a \in \{+1, -1\}$ gives $s_x = m_a$. When $|\mathcal{P}_a| = 2$, the physical qubit encodes the parity $q_x \oplus q_y$, and its measurement outcome satisfies $m_a = s_x \cdot s_y$ in the spin representation ($s = +1 \leftrightarrow 0$, $s = -1 \leftrightarrow 1$). If one of the two logical spins, say $s_x$, has already been decoded from a previously processed physical qubit, the other can be recovered as $s_y = m_a \cdot s_x$.

The choice of which parity map to use and the order in which physical qubits are processed depend on the parity of the number of QAOA layers $p$. Since the PTC 
protocol alternates the direction of the encoding between consecutive layers (forward for even layers, backward for odd layers), the final parity map 
corresponds to:

\begin{itemize}
    \item If $p$ is even: the initial parity map $\mathcal{P}^0$ (time step 
    $t = 0$).
    \item If $p$ is odd: the final parity map $\mathcal{P}^{D-1}$ (time step 
    $t = D - 1$, where $D$ is the encoding depth).
\end{itemize}

In both cases, the traversal order ensures that single-qubit parities 
($|\mathcal{P}_a| = 1$) are encountered before their dependent two-qubit parities, enabling sequential resolution of all logical spins. The full procedure is given in Algorithm~\ref{alg:ptc_decode}.

\begin{algorithm}[t]
\caption{PTC Parity Decoding}\label{alg:ptc_decode}
\KwIn{Measurement bitstring $\mathbf{b} = (b_0, b_1, \ldots, b_{N_q-1})$,
      number of QAOA layers $p$,
      PTC parity structure $\{\mathcal{P}^t\}_{t=0}^{D-1}$}
\KwOut{Decoded logical bitstring $\mathbf{x} = (x_0, x_1, \ldots, x_{N_q-1})$}

\BlankLine
\tcp{Convert measurement to spin representation}
\For{$i = 0$ \KwTo $N_q - 1$}{
    $m_i \gets (+1)^{1 - b_i}$  \tcp*{$0 \mapsto +1$, $1 \mapsto -1$}
}

\BlankLine
\tcp{Select parity map and traversal order based on layer parity}
\eIf{$p \bmod 2 = 0$}{
    $\mathcal{P} \gets \mathcal{P}^0$ \tcp*{Initial parity map}
    $\mathcal{Q} \gets (0, 1, \ldots, N_q - 1)$ \tcp*{Forward order}
}{
    $\mathcal{P} \gets \mathcal{P}^{D-1}$ \tcp*{Final parity map}
    $\mathcal{Q} \gets (D{-}1, D{-}2, \ldots, 0,\; D, D{+}1, \ldots, N_q - 1)$ \tcp*{Reverse-then-forward}
}

\BlankLine
\tcp{Initialize logical spin array}
$s_j \gets 0$ for all $j \in \{0, 1, \ldots, N_q - 1\}$\;

\BlankLine
\tcp{Sequential parity resolution}
\ForEach{$a \in \mathcal{Q}$}{
    $(x, y, \ldots) \gets \mathcal{P}_a$\;
    \uIf{$|\mathcal{P}_a| = 1$ \tcp*{Single logical qubit}}{
        $s_x \gets m_a$\;
    }
    \ElseIf{$|\mathcal{P}(i)| = 2$ \tcp*{Parity of two logical qubits}}{
        \uIf{$s_x \neq 0$}{
            $s_y \gets m_a \cdot s_x$ 
        }
        \ElseIf{$s_y \neq 0$}{
            $s_x \gets m_a \cdot s_y$\;
        }
    }
}

\BlankLine
\tcp{Convert spins back to binary}
\For{$j = 0$ \KwTo $N_q - 1$}{
    $x_j \gets \frac{1 - s_j}{2}$ \tcp*{$+1 \mapsto 0$, $-1 \mapsto 1$}
}

\Return $\mathbf{x}$\;
\end{algorithm}

\section{Results}\label{Sec:Results}

In this section, results comparing PTC+SA, SWAP+SA, and the methods of Sec.~\ref{Sec:transpilation} for transpiling a QAOA layer for WMC problems of different $E_d$ are presented. In addition, how depolarizing noise affects the LR-QAOA solution of a WMC problem is shown, and experimental results on ibm\_fez for LR-QAOA fully connected WMC problems using the SWAP and PTC techniques are presented.

\subsection{Transpilation results}

Figure~\ref{fig:transpilation} shows the resources needed to implement a layer of QAOA using different methods. In Fig.~\ref{fig:transpilation}(a), the number of two-qubit gates over the minimum number of two-qubit gates is shown for problems involving $N_q = 20$. At small connectivity in the graph, i.e., \(E_d = 0.211\), the $N_g$ of PTC+SA and SWAP+SA require more than 1.5 times the gates needed by Qiskit-T  and Qiskit-AI. This results from the fact that the transpilers exploit the heavy-hex layout and use ancillary qubits to reduce \(N_g\). The trend of the transpilers giving a lower \(N_g\) changes at $E_d \approx 0.35$, where the PTC already gives a similar two-qubit gate count to the Qiskit transpiler. The TKET transpiler requires a similar ratio $N_g/N_g^{min}$ throughout the \(E_d\), which is above the other transpilers. As Qiskit-P is based on the SWAP network strategy, the observed gap between Qiskit-P and SWAP+SA quantifies the additional improvement introduced by the SA step.

In Fig.~\ref{fig:transpilation}(b), the circuit depth over the minimum possible depth is presented for the different methods. An improvement in depth for the SWAP+SA and PTC+SA compared with the transpilers is seen for all $E_d$. Therefore, even when the QAOA circuit is sparsely connected and certain transpilers achieve a lower \(N_g\), this often comes at the expense of a significant increase in circuit depth. Fig.~\ref{fig:transpilation}(c)–(e), which show a QAOA layer for PTC+SA, SWAP+SA, and Qiskit-T (from left to right), highlights the compactness achieved by PTC+SA and SWAP+SA while Qiskit-T uses fewer \(N_g\) at the expense of increased circuit depth.

Table~\ref{Table:comparison} presents a comparison of various transpilation methods across different edge densities ($E_d$) and qubit counts ($N_q = 20$, $60$, and $120$). For each setting, the number of two-qubit gates ($N_g$), circuit depth ($d$), and runtime in seconds ($t$) are reported. The methods include the minimum \(N_g\) and \(d\) for a QPU with a fully connected layout (FC) and access to ZZ gates. PTC+SA consistently achieves the lowest circuit depths across all configurations, demonstrating its effectiveness in producing compact circuits. While Qiskit-T often minimizes the two-qubit gate count, this typically comes at the expense of significantly higher depth, particularly as $E_d$ increases. SWAP+SA shows similar depth performance to PTC+SA with a moderate increase in $N_g$. Qiskit-AI and TKET generally incur higher depth or gate counts, and therefore, only their analysis for the 20-qubit problems is included. Runtime remains low for Qiskit-T and Qiskit-P, and considerable time is needed for PTC+SA and SWAP+SA at low \(E_d\), but it decreases quickly as \(E_d\) grows. Qiskit-AI incurs longer times, mainly due to the overhead of sending the circuit.

\begin{table*}[ht]
\centering
\caption{Comparison of methods across various $E_d$ levels and qubit counts.}
\renewcommand{\arraystretch}{1.2}
\setlength{\tabcolsep}{4pt}
\begin{tabular}{|c|c||ccc|ccc|ccc|ccc|ccc|}
\hline
$N_q$ & Method& $N_{2q}$ & $d$ & $t$ (s)& $N_{2q}$ & $d$ & $t$ (s)& $N_{2q}$ & $d$ & $t$ (s)& $N_{2q}$ & $d$ & $t$ (s)& $N_{2q}$ & $d$ & $t$ (s) \\
\hline \hline
\multirow{8}{*}{20} & & \multicolumn{3}{c|}{${E_d} = 0.211$} & \multicolumn{3}{c|}{${E_d} = 0.298$} & \multicolumn{3}{c|}{${E_d} = 0.386$} & \multicolumn{3}{c|}{${E_d} = 0.649$} & \multicolumn{3}{c|}{${E_d} = 1.0$} \\
\cline{3-17}
& FC & 41 & 12 & - & 57 & 15 & - & 74 & 18 & - & 124 & 22 & - & 190 & 22 & - \\
& Qiskit-P & 519& 72& 0.06& 523& 73& 0.17& 527& 77& 0.09& 537& 81& 0.10& 551& 81& 0.13\\
& Qiskit-T & 185& 65& 0.10& \underline{\textbf{282}}& 91& 0.11& 362& 131& 0.13& 601& 159& 0.16& 739& 183& 0.36\\
& Qiskit-AI & \underline{\textbf{173}}& 87& 21.54& 312& 142& 22.65& 416& 189& 23.04& 640& 282& 24.39& 709& 195& 22.26\\
& HE-TKET & 302& 97& 1.96& 441& 159& 62.40& 524& 195& 6.29& 835& 347& 69.14& 1191& 420& 66.99\\
& PTC+SA & 283& \underline{\textbf{44}}& 2.02& 302& \underline{\textbf{48}}& 2.44& \underline{\textbf{340}}& \underline{\textbf{54}}& 2.35& \underline{\textbf{378}}& \underline{\textbf{61}}& 1.86& \underline{\textbf{399}}& \underline{\textbf{66}}& 1.47\\
& SWAP+SA & 318& 44& 1.20& 380& 54& 1.25& 407& 60& 1.40& 469& 69& 1.12& 532& 77& 0.89\\
\hline
\multirow{6}{*}{60} & & \multicolumn{3}{c|}{${E_d} = 0.068$} & \multicolumn{3}{c|}{${E_d} = 0.171$} & \multicolumn{3}{c|}{${E_d} = 0.275$} & \multicolumn{3}{c|}{${E_d} = 0.586$} & \multicolumn{3}{c|}{${E_d} = 1.0$} \\
\cline{3-17}
& FC & 121 & 15 & - & 303 & 29 & - & 487 & 42 & - & 1038 & 59 & - & 1770 & 62 & - \\
& Qiskit-P & 5139& 198& 0.40& 5153& 211& 0.36& 5167& 224& 0.58& 5199& 239& 1.08& 5251& 241& 1.94\\
& Qiskit-T & \underline{\textbf{922}}& 178& 0.24& 2853& 543& 0.45& 4228& 700& 0.65& 6723& 1068& 1.22& 7530& 896& 1.59\\
& PTC+SA & 2004& \underline{\textbf{84}}& 4.71& \underline{\textbf{2771}}& \underline{\textbf{125}}& 5.34& \underline{\textbf{3125}}& \underline{\textbf{148}}& 6.26& \underline{\textbf{3479}}& \underline{\textbf{176}}& 7.13& \underline{\textbf{3599}}& \underline{\textbf{186}}& 3.25\\
& SWAP+SA & 2667& 108& 2.27& 3989& 168& 2.57& 4437& 191& 3.74& 4894& 224& 3.03& 5192& 237& 1.72\\
\hline
\multirow{6}{*}{120} & & \multicolumn{3}{c|}{${E_d} = 0.034$} & \multicolumn{3}{c|}{${E_d} = 0.141$} & \multicolumn{3}{c|}{${E_d} = 0.248$} & \multicolumn{3}{c|}{${E_d} = 0.57$} & \multicolumn{3}{c|}{${E_d} = 1.0$} \\
\cline{3-17}
& FC & 243 & 15 & - & 1007 & 50 & - & 1771 & 76 & - & 4070 & 117 & - & 7140 & 122 & - \\
& Qiskit-P & 21075& 384& 0.93& 21095& 418& 1.58& 21119& 439& 2.55& 21201& 476& 8.81& 21301& 481& 25.38\\
& Qiskit-T & \underline{\textbf{2985}}& 378& 0.76& 12926& 1786& 3.06& 18769& 2417& 4.12& 29642& 3536& 6.98& 34371& 3123& 9.02\\
& PTC+SA & 7852& \underline{\textbf{154}}& 10.05& \underline{\textbf{12374}}& \underline{\textbf{266}}& 10.24& \underline{\textbf{13326}}& \underline{\textbf{304}}& 13.01& \underline{\textbf{14159}}& \underline{\textbf{354}}& 15.22& \underline{\textbf{14399}}& \underline{\textbf{366}}& 6.59\\
& SWAP+SA & 10897& 205& 4.69& 17519& 354& 5.31& 19300& 402& 5.58& 20576& 461& 6.88& 21182& 477& 3.19\\
\hline
\end{tabular}
\label{Table:comparison}
\end{table*}

\begin{figure*}[!tbh]
    \centering
    \includegraphics[width=1\linewidth]{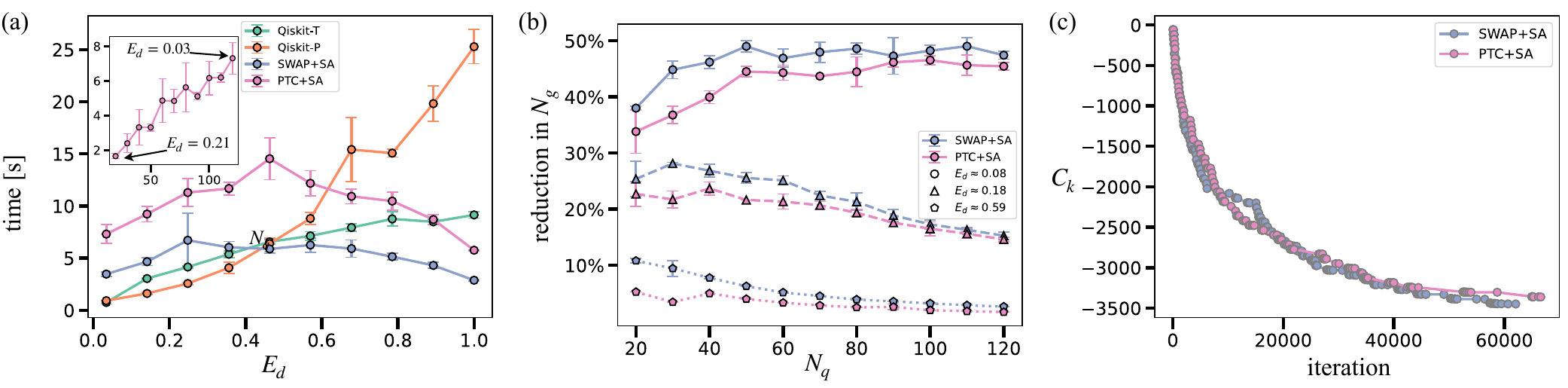}
    \caption{(a) Runtime vs. $E_d$ for the different transpilers to find a circuit that encodes a layer of QAOA for 120-qubit random problems. The error bars in all three plots represent the mean and standard deviation computed over three different randomly generated graphs. (b) Reduction in two-qubit gates compared to the PTC and SWAP default methods for 3 different \(E_d\) values. (c) SA \(C_k\) function vs. iteration for PTC and SWAP.}
    \label{fig:time}
\end{figure*}

Figure~\ref{fig:time}(a) shows the transpilation time needed to find the circuit representing the 120-qubit problems. Qiskit-T and Qiskit-P take short times at low \(E_d\) compared to PTC+SA and SWAP+SA, but the trend quickly changes as \(E_d\) grows. The inset plot shows the time taken at the minimum \(E_d\) for each problem size for the PTC+SA transpilation. For small values of \(E_d < 0.14\), the use of PTC+SA offers no significant benefit over Qiskit-T in terms of execution time or the number of gates  \(N_g\). However, beyond this threshold, PTC+SA yields a reduction in both \(N_g\) and \(d\), while maintaining similar execution times. The largest runtimes for the PTC+SA and SWAP+SA methods, which increase from approximately 2 seconds at 20 qubits to around 10 seconds at 120 qubits, remain practical for optimizing large-scale quantum circuits.

Fig.~\ref{fig:time}(b) shows the reduction in \(N_g\) thanks to the use of SA compared to the raw circuit produced by PTC and SWAP strategies. The reduction is more notable as the QAOA circuit is more sparsely connected, reaching a point where there is a reduction of approximately \(50\%\) for large \(N_q\). Fig.~\ref{fig:time}(c) shows the cost function evaluation at each iteration of the SA function for PTC and SWAP by the number of iterations used for the \(N_q=120\) and \(E_d=0.03\). 

\begin{figure}[!tbh]
    \centering
    \includegraphics[width=1\linewidth]{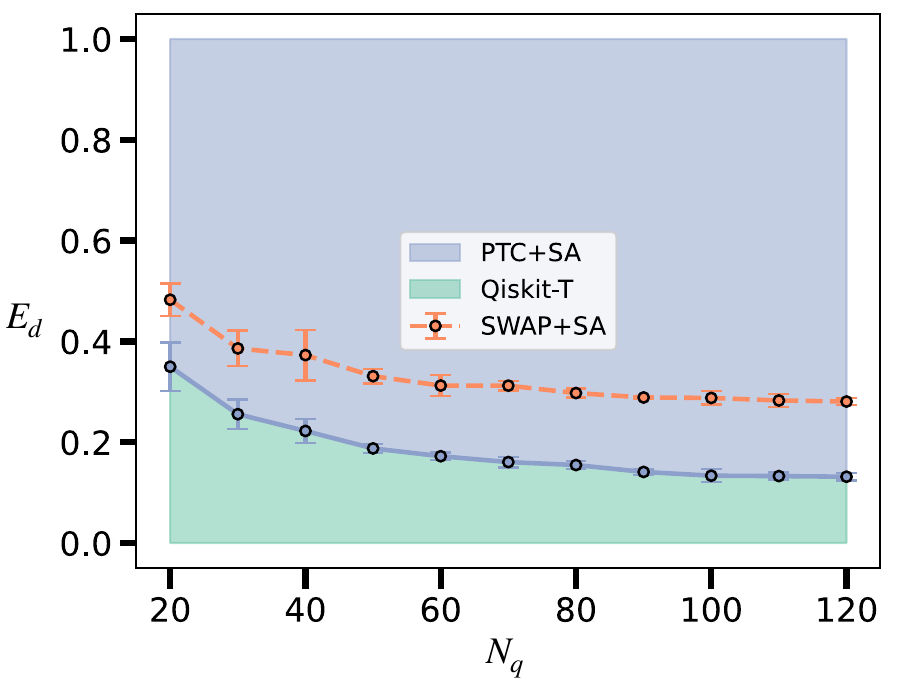}
    \caption{Graph density vs. number of qubits. The blue region indicates where PTC requires fewer two-qubit gates than the Qiskit transpiler level 3. At the largest size, 120 qubits and $E_d > 0.13$, PTC requires fewer two-qubit gates than Qiskit. The orange dashed line divides the region where the SWAP+SA strategy gives a lower two-qubit count than Qiskit-T. The error bars in the markers are the standard deviation for three random graphs at each $E_d$.}
    \label{fig:PTCgain}
\end{figure}

Figure~\ref{fig:PTCgain} shows the values of \(E_d\) for which the PTC+SA and SWAP+SA encodings result in a lower number of two-qubit gates compared to Qiskit-T. The blue region represents the case when the PTC+SA requires fewer two-qubit gates, and above the dashed line when the SWAP+SA needs a lower \(N_g\) than Qiskit-T. For example, when \(N_q = 20\), a gate count reduction with PTC+SA is observed only when the circuit's connectivity exceeds \(E_d = 0.35\). However, as the system size increases, this threshold decreases. For \(N_q = 80\), the required connectivity drops to \(E_d = 0.15\), and for \(N_q = 120\), it decreases further to \(E_d = 0.13\). In the SWAP+SA case, at 20 qubits, the mean crossing point is at \(E_d=0.48\), and as the system size increases, the mean crossing points gradually converge at approximately \(E_d=0.28\) for 120 qubits. This is particularly relevant since PTC+SA and SWAP+SA encodings significantly reduce the depth and do not use additional ancilla qubits.

\begin{figure}[!tbh]
    \centering
    \includegraphics[width=1\linewidth]{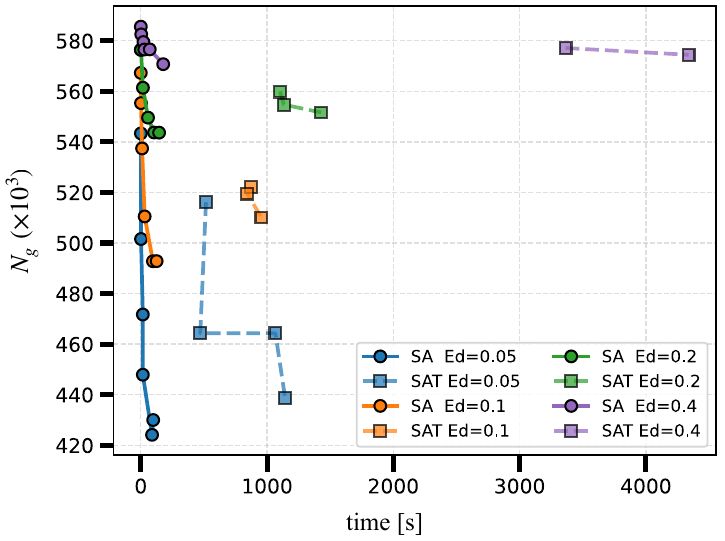}
    \caption{SA and SATMapper (SAT) comparison of the number of two-qubit gates versus execution time for a 200-qubit problem with $p=10$ and $E_d \in \{0.05, 0.1, 0.2, 0.4\}$. Different markers indicate increasing runtime. For the SA approach, the iterations considered $\in \{10^3, 10^4, 10^5, 2\times10^5, 5\times10^5, 10^6\}$, while for SAT the timeout limits are $\in \{0.1, 1, 2, 5, 10, 60\}$ seconds.}
    \label{fig:SATcomparison}
\end{figure}

For a 200-qubit QAOA circuit on a 1D-chain topology ($p=10$), Fig.~\ref{fig:SATcomparison} shows that SA consistently outperforms SATMapper in practical runtime by orders of magnitude: SA converges to good solutions in seconds, whereas the SAT solver requires hundreds to thousands of seconds of wall-clock time regardless of its timeout limit per iteration. Despite this speed improvement, SA approaches SAT-quality solutions at sufficient iteration counts. For example, at an edge density of $E_d=0.05$, SA with $10^6$ iterations yields 430\,k two-qubit gates compared to 439\,k for SAT (60\,s timeout), a difference of less than 2\% in 98\,s compared to 1143\,s for SATMapper. For denser graphs ($E_d=0.4$), both methods converge to essentially the same gate count ($\sim$572\,k), while SA requires only $\sim$180\,s versus over one hour for SAT. Furthermore, SA exhibits a diminishing-returns profile with respect to iteration budget. Thus, for sparse graphs, the largest reduction in two-qubit gate count occurs before $2\times10^5$ iterations, suggesting that a moderate iteration budget (e.g., $10^5$) offers a favorable quality-to-runtime trade-off. Additionally, for small timeouts (0.1 and 1\,s), the SATMapper solver may fail to find a valid solution. This explains the missing markers in the plot, whereas SA always produces a valid solution since it starts from a feasible mapping.

\subsection{Noise simulation}

\begin{figure*}[!tbh]
    \centering
    \includegraphics[width=0.75\linewidth]{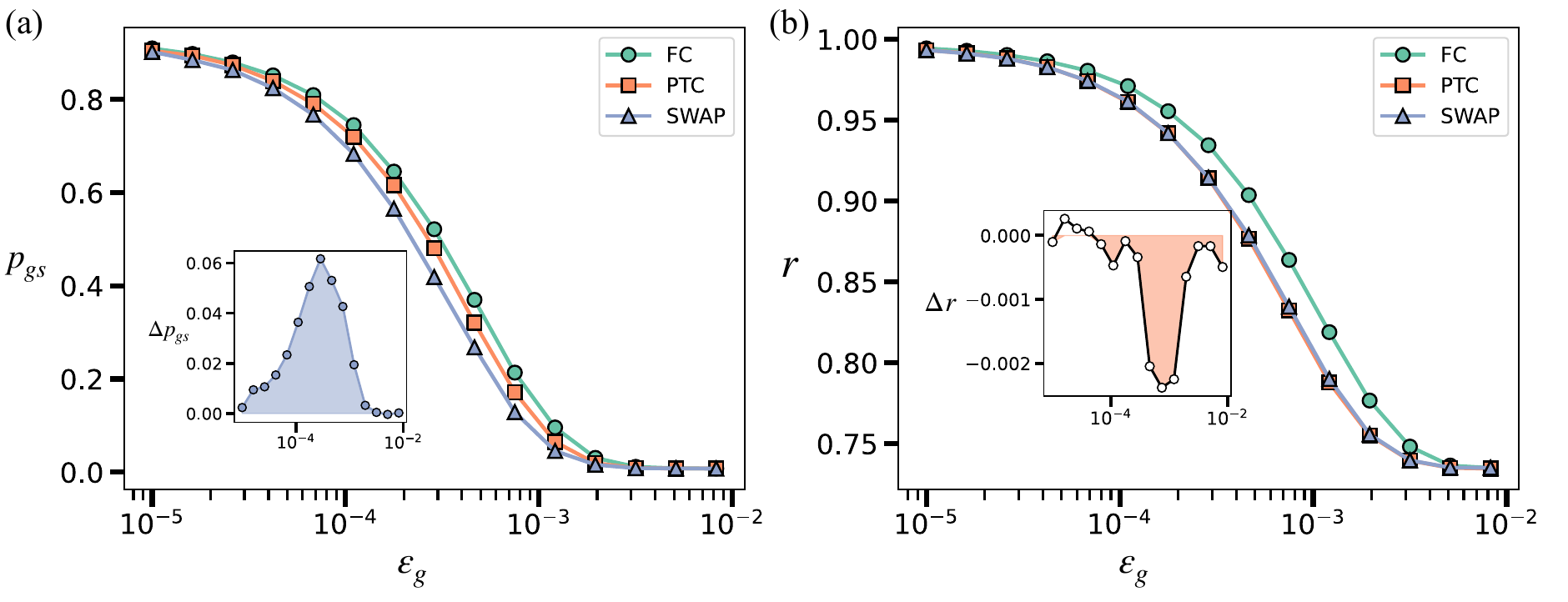}
    \caption{Depolarizing noise simulation of an 8-qubit WMC using LR-QAOA with $p=50$. (a) Success probability vs. depolarizing noise for a fully connected (FC) device and PTC and SWAP encodings in a 1D-chain. The inset graph shows the difference in $p_{gs}$ of the PTC and SWAP. (b) Approximation ratio vs. the depolarizing noise error for the three models. The inset graph shows the difference between PTC and SWAP; a negative value indicates an improvement of PTC over SWAP.}
    \label{fig:noisesim}
\end{figure*}

Figure~\ref{fig:noisesim} shows the impact of depolarizing noise on the two-qubit-gate depolarizing error, $\varepsilon_g$, relative to the performance of LR-QAOA using different problem encodings for an 8-qubit WMC problem with $p=50$. Results are presented for the simulation of a device with full connectivity between its qubits (FC), the PTC, and SWAP in a 1D chain of qubits. Fig.~\ref{fig:noisesim}(a) shows the success probability at different depolarizing noise strengths in the two-qubit gates. In this case, the method that decays faster as the error increases is the SWAP method, which is mainly explained by the fact that it requires $N_g = 3,850$ while the PTC requires $N_g = 3,100$, and the FC requires $N_g = 2,800$ CNOT gates. The inset shows $\Delta p_{gs} = p_{gs}^{\text{PTC}} - p_{gs}^{\text{SWAP}}$, highlighting the improvement of the PTC encoding over the SWAP strategy as noise increases. Initially, the PTC encoding gives a higher success probability, indicating better performance. However, beyond a certain noise threshold, the improvement diminishes, and no benefit is observed under stronger noise conditions. At the peak, the PTC gives 5.65\% more optimal solutions than the SWAP strategy for an $\varepsilon_g = 4.6 \times 10^{-4}$. 

In the case of the approximation ratio, Fig.~\ref{fig:noisesim}(b), the difference between PTC and SWAP is small, which is somewhat unexpected, as the PTC approach requires 750 fewer two-qubit gates compared to the SWAP method. The inset plot shows the difference, $\Delta r = r^{\text{PTC}} - r^{\text{SWAP}}$. It seems that maintaining logical consistency in the averaged outcomes is more difficult with the PTC encoding. The shaded region represents where the PTC encoding gets worse average energy than the SWAP. However, this difference is too small to draw any conclusion about whether either of the two encodings is better. Note that in the same noise range, the $p_{gs}$ improves in the PTC encoding.

\subsection{Real QPU performance}

\begin{figure*}[!tbh]
    \centering
    \includegraphics[width=0.75\linewidth]{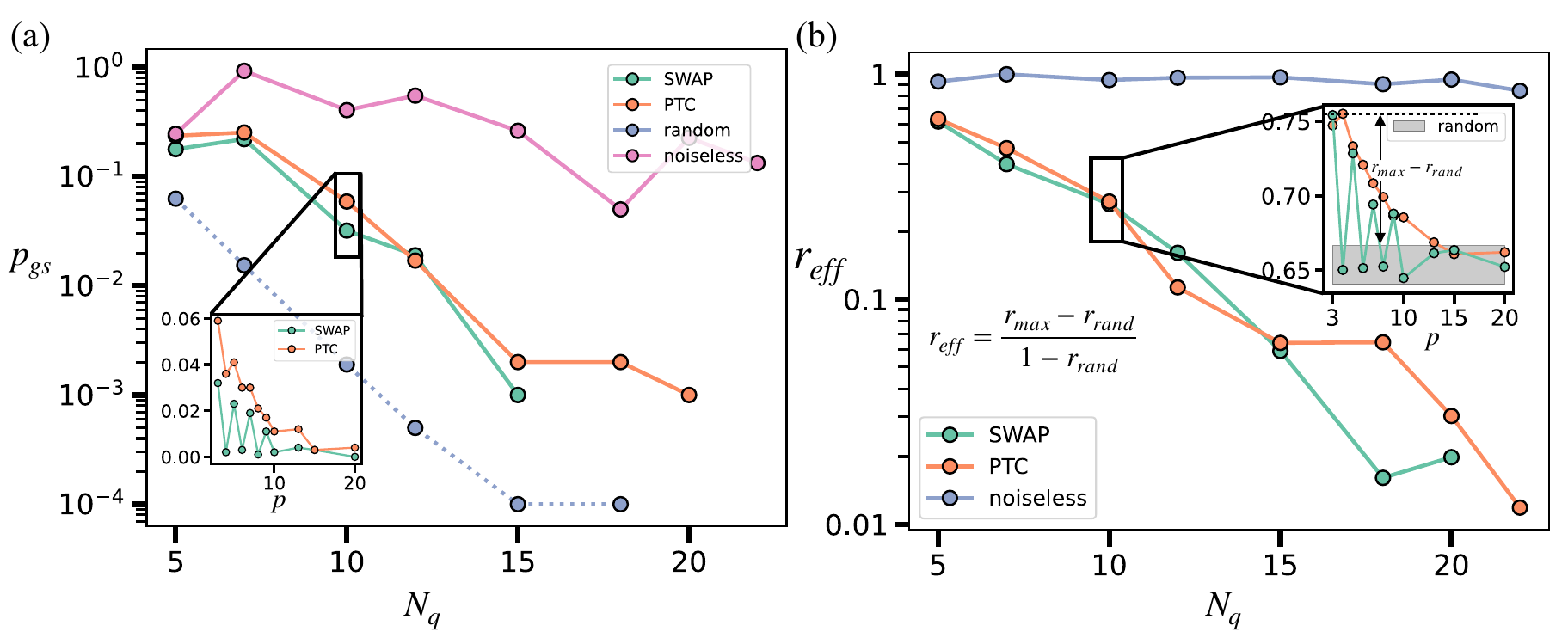}
    \caption{LR-QAOA experiments on ibm\_fez for WMC problems ranging from 5 to 22 qubits using PTC and SWAP encodings and comparing them against a random sampler and a noiseless simulation. (a) Best $p_{gs}$ vs. the number of qubits for the SWAP, PTC, a noiseless simulation, and a random sampler. The inset graph shows the probability of success, $p_{gs}$ versus the number of layers for a 10-qubit problem. (b) Effective approximation ratio, $r_{\text{eff}}$, vs. the number of qubits. The inset graph shows $r$ vs the number of layers for a 10-qubit problem.  }
    \label{fig:pgs_QPU}
\end{figure*}

Figure~\ref{fig:pgs_QPU}(a) shows the performance of SWAP and PTC encodings for a fully connected weighted MaxCut problem at different problem sizes and depths. Fig.~\ref{fig:pgs_QPU}(a) shows the maximum success probability for a given number of qubits using the SWAP and PTC encodings. This probability is compared with the random guessing probability and the LR-QAOA noiseless simulation, and an at least an order-of-magnitude larger $p_{gs}$ is obtained using the PTC encoding. The optimal solution is even found at 20 qubits with 1000 samples, while using the SWAP encoding, the optimal solution is only found up to 15 qubits. In the case of $N_q=20$, using PTC, the optimal solution is found at $p=4$, which requires $N_g = 1,195$. 

In the inset of Fig.~\ref{fig:pgs_QPU}(a), there is an improvement in $p_{gs}$ with a peak at $p=3$. At this point, the PTC probability of success is almost twice as large as in the SWAP case. This improvement can be attributed mainly to the reduction in the two-qubit gate number, where it passes from 405 in SWAP to 295 in PTC at $p=3$. 

Figure~\ref{fig:pgs_QPU}(b) shows the approximation ratio behavior for WMC problems using LR-QAOA and the PTC and SWAP encodings on ibm\_fez. Figure~\ref{fig:pgs_QPU}(b) shows the effective approximation ratio (inset formula), which is greater than zero if the encoding is better than the upper part of the random region. Experiments with up to 25 qubits were implemented, but results better than those of the random sampler were only found for 20 qubits using SWAP and 22 qubits using PTC. Thus, employing PTC, extends the useful LR-QAOA experiment by two qubits, which requires $N_g = 1,449$ at $p=3$. 

The inset in Fig.~\ref{fig:pgs_QPU}(b) shows the approximation ratio versus the number of LR-QAOA layers. In this case, the SWAP encoding has an oscillating behavior that might indicate that there is a destructive error when the number of LR-QAOA layers is even. The random region is the approximation ratio interval of confidence where 1,000 sampled bitstrings can be explained by a random sampler (see Appendix A-2 in \cite{montanezbarrera2025}). Therefore, when the approximation ratio from a QPU's bitstrings fails this test, it means that the LR-QAOA logic in the QPU is completely suppressed. If only the odd points of the SWAP method and the PTC points are taken, the conclusion is similar to the simulation using depolarizing noise. Both methods give similar results, even if the PTC method requires fewer two-qubit gates.

\begin{figure}[!tbh]
    \centering
    \includegraphics[width=1\linewidth]{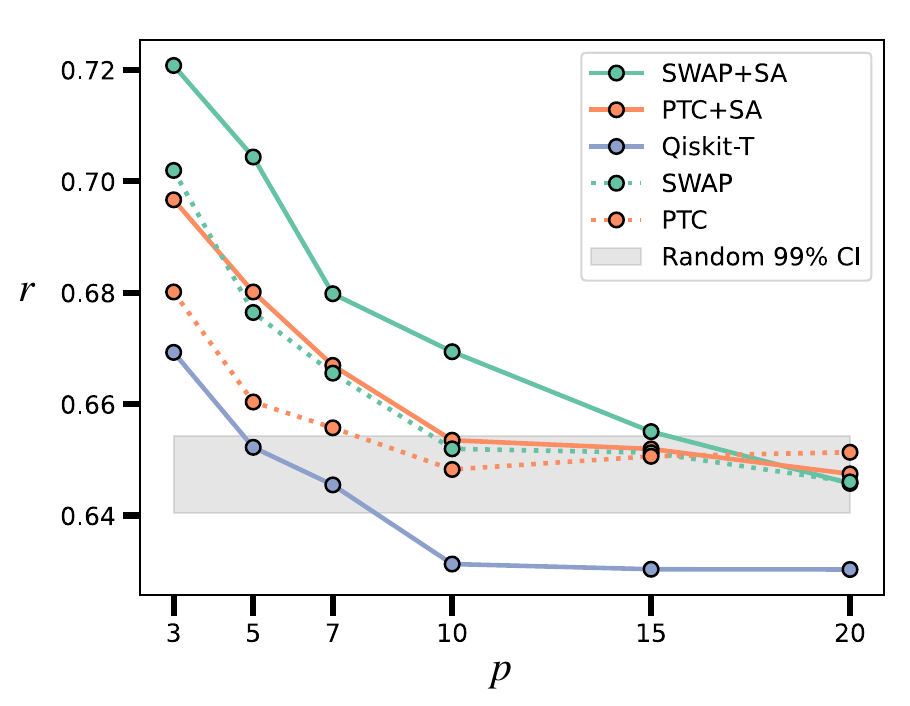}
    \caption{LR-QAOA experiments on ibm\_kingston for the \(E_d=0.298\) 20-qubit WMC problem of Table \ref{tab:ed_comparison} using PTC, SWAP encodings both with SA to reduce gate count (PTC+SA, SWAP+SA) and without it (PTC, SWAP), as well as Qiskit-T; comparison is made against a random sampler.}
    \label{fig:ibm_kingston}
\end{figure}

Figure~\ref{fig:ibm_kingston} shows the experimental comparison of transpilation methods on ibm\_kingston at different LR-QAOA depths. SWAP+SA achieves the best results, reaching $r = 0.7208$ at $p = 3$ (a $20.83\%$ improvement over random), while PTC+SA's improvement is only $13.99\%$ over random. In contrast, Qiskit-T's standard transpilation degrades quickly, approaching the random baseline at low depth.

It is somewhat unexpected that the SWAP without SA (SWAP) shows better performance than PTC+SA at $p = 3$, even though SWAP requires significantly more two-qubit gates (1596 vs. 963). This suggests that performance is not determined only by the number of two-qubit gates, but also by the encoding itself. A possible explanation is that encoding the logic of pairs of qubits into single qubits, as is done in PTC, may be more sensitive to noise, since errors affecting a single qubit can impact multiple logical correlations. This behavior is also observed in fully connected experiments, where SWAP encoding can sometimes achieve better approximation ratios despite its larger gate count. In this case, the effect appears more pronounced, possibly due to the reduced number of gates, highlighting that the trade-off between gate count and noise sensitivity depends strongly on the encoding structure.

An analysis of the output bit distributions explains why Qiskit-T even goes below the random baseline. Qiskit-T exhibits a strong Hamming weight bias, which is $5.84\times$ worse than SWAP and $16.45\times$ worse than PTC. This bias leads to samples skewed toward zeros (a mean Hamming weight of $\sim 8.5$ while ideally $\sim 10$ because of the $\mathbb{Z}_2$ symmetry of the problem) that might indicate a relaxation phenomenon that can be attributed to the large circuit depth. The net effect is observed by an individual qubit's bias towards the classical ground state $|0\rangle^{\otimes n}$. In contrast, SWAP and PTC maintain distributions with low bias (bias $<0.25$ bits) across all depths, preserving quantum correlations through more structured qubit routing and without ancilla usage.

\section{Conclusions}\label{Sec:Conclusions}
In this work, we evaluate the use of PTC and SWAP encodings to represent QAOA circuits with arbitrary connectivity using a 1D chain of qubits. Additionally, we introduce a simulated annealing technique to reduce the number of two-qubit gates in non-fully connected circuits. While the PTC or SWAP encoding is primarily designed for fully connected circuits, we demonstrate that the combination of PTC or SWAP with simulated annealing, PTC+SA and SWAP+SA, yields a further reduction in two-qubit gate count. 

The PTC+SA and SWAP+SA methods give a reduction in two-qubit gates and depth compared to Qiskit’s transpiler at optimization level 3, when the circuit’s edge density \(E_d\) exceeds a threshold. For example, in a 20-qubit circuit, this threshold is \(E_d=0.35\) for PTC+SA, while in a 120-qubit circuit, it decreases to \(E_d=0.13\). The proposed SA-based optimization algorithm reduces this threshold by relabeling qubits such that the two-qubit pairs appearing at the end of the PTC or SWAP protocols correspond to interactions not required in the original circuit, and can therefore be removed. While other metaheuristics such as genetic algorithms or tabu search could also be considered, SA offers a simple and effective approach for this problem. A systematic comparison with alternative optimization strategies is left for future work.

We compare our SA method against the recently implemented SATMapper strategy in Qiskit \cite{qopt_best_practices} for the subgraph isomorphism problem in QAOA circuits with 200 qubits at $p = 10$ across different edge densities, and observe significant reductions in runtime and two-qubit gate count.

Additionally, we simulate an 8-qubit fully connected LR-QAOA circuit with \(p=50\) using both the PTC and SWAP encodings under a depolarizing noise model. The results show a slight improvement in the success probability when using the PTC encoding compared to the SWAP encoding. However, in terms of the approximation ratio, both methods perform similarly, with no appreciable advantage observed for PTC. Overall, under equivalent noise conditions, the performance of both PTC and SWAP encodings remains comparable to that of a device with a fully connected layout, despite the additional overhead required to represent the fully connected circuit on a 1D chain of qubits.

The PTC encoding of an LR-QAOA protocol used in quantum benchmarking \cite{montanezbarrera2025} is tested. Some improvement is found compared to the SWAP method in terms of the quality of the solutions. For instance, under identical conditions, optimal solutions are obtained using PTC for problem sizes up to $N_q=20$, whereas the SWAP method yields optimal solutions only up to $N_q=15$. In terms of the approximation ratio quality, both methods give similar results, which agree with the noise simulation using depolarizing noise. However, it is found that the LR-QAOA implementation on ibm\_fez remains distinguishable from a random sampler for problem sizes up to 22 qubits when using the PTC method, whereas this distinction holds only up to 20 qubits when using the SWAP strategy.

Finally, experiments on ibm\_kingston for a graph with $E_d = 0.298$ show that SWAP+SA performs best (20.83\% improvement over random), followed by PTC+SA, while Qiskit-T performs worse as depth increases and eventually reaches or drops below random performance. This results from a strong Hamming weight bias caused by relaxation toward $\left|0\right\rangle$ of individual qubits in Qiskit-T. In contrast, simpler routing methods such as SWAP better preserve the quantum signal. Although PTC uses fewer gates, it still performs worse than SWAP in non-fully connected experiments. This may be because encoding two-qubit operations into single qubits leads to faster error accumulation, but further analysis, which is beyond the scope of the present work, is needed.

\section*{Data Availability}
All problem instances and results analyzed in this study are available at: https://github.com/alejomonbar/QAOA-efficient-circuit-transpilation

\section*{Acknowledgment}

J. A. Montanez-Barrera and Y. Ji acknowledge support from the German Federal Ministry of Education and Research (BMBF), the funding program Quantum Technologies - from basic research to market, project QSolid (Grant No. 13N16149). We acknowledge the use of IBM Quantum services for this work. The views expressed are those of the authors and do not reflect the official policy or position of IBM or the IBM Quantum team.
Y. Ji acknowledges support from the project ``Quantum-based
Energy Grids (QuGrids)”, funded by the program ``Profilbildung 2022”, an initiative of the Ministry of Culture and Science of the State of North Rhine-Westphalia.

\clearpage
\bibliography{References}

\end{document}